\documentclass[aps,prl,reprint,groupedaddress,showpacs,amsmath, amssymb, floatfix]{revtex4-1}
\usepackage{epstopdf}
\usepackage{graphicx}
\usepackage{color}
\usepackage{latexsym}
\usepackage{amssymb}
\usepackage{bm}
\usepackage{dcolumn}

\begin{document}

\title{Corrections to Higgs Mode Masses in Superfluid $^3$He from Acoustic Spectroscopy}

\author{M. D. Nguyen}
\email{mannguyen2019@u.northwestern.edu}
\author{A.M Zimmerman}
\author{W.P. Halperin}
\email{w-halperin@northwestern.edu}
\affiliation{Department of Physics and Astronomy, Northwestern University \\Evanston, IL 60208, USA}

\date{\today}

\begin{abstract}
Superfluid $^3$He has a rich spectrum of collective modes with both massive and massless excitations. The masses of these modes can be precisely measured using acoustic spectroscopy and fit to theoretical models. Prior comparisons of the experimental results with theory did not include strong-coupling effects beyond the weak-coupling-plus BCS model, so-called non-trivial strong-coupling corrections. In this work we utilize recent strong-coupling calculations to determine the Higgs masses and find consistency between experiments that relate them to a sub-dominant $f$-wave pairing strength. 
\end{abstract}

\maketitle

\section{I. Introduction}
Collective modes are integral to the understanding of many-body physics because they reflect the broken symmetries for condensed phases and encode the dynamical response of a system to external forces. The dynamics of excitations in superfluid $^3$He depend upon whether or not an energy gap exists in their corresponding energy-momentum relation, which is analogous to the mass of elementary particles \cite{nam.61}.
In a 1985 paper, Nambu observed that the masses of bosonic collective modes of superfluid $^3$He-B, including the analog of the Higgs boson, are related to the mass of the fermionic excitations through a sum rule \cite{nam.85}. Further, he speculated on the possibility that there is a hidden supersymmetry associated with the class of field theories of the Bardeen-Cooper-Schrieffer (BCS) type, including superfluid $^3$He. Following the discovery of the Higgs boson, there has been renewed interest in the Nambu sum rule and its connection to $^3$He-B \cite{vol.14}, including a recent report showing that the Nambu sum rule is not exact \cite{sau.17}. The sum rule is violated due to mass renormalization from the polarization of the underlying fermionic vacuum as well as strong-coupling corrections to the BCS theory \cite{sau.17,sau.81a}.

Higgs modes have been studied in several other systems including superconductors NbSe$_2$ using Raman scattering \cite{soo.80,lit.82} and Nb$_{1-x}$Ti$_{x}$N with THz excitation \cite{mat.14}. The results reported here are based upon acoustic spectroscopy of two Higgs modes in superfluid $^3$He-B \cite{mas.80,gia.80,fra.89,dav.06}. We analyze these measurements using the theoretical calculations of the mass corrections reported in Ref. \cite{sau.17} to determine fundamental interactions of the system.

Superfluidity in $^3$He arises when the quasiparticles of the normal Fermi liquid condense into a $p$-wave, spin-triplet superfluid ($L = S = 1$) that can be understood from the BCS pairing theory for superconductivity~\cite{vol.90}. The superfluid state breaks not only the U(1) symmetry of the normal liquid but also reduces the separate orbital and spin rotation symmetries to a combined SO(3)$_{L+S}$ residual symmetry in the B-phase. This complex pattern of symmetry breaking gives rise to 18 collective modes that are labeled by two quantum numbers \footnote{The third quantum number, $m$, corresponding to the angular momentum projection counts the degeneracy in zero magnetic field}; the total angular momentum, $J\,\,(= 0,1,2)$,  and the particle-hole conjugation parity, $c$ ($+, -$)\cite{hal.90,sau.17}.
Four of these modes are massless, Nambu-Goldstone bosons while 14 have a gap in their energy dispersion, corresponding to the Higgs masses \cite{sau.17,zav.16}. 

The bare masses were calculated in the weak-coupling BCS theory by several authors \cite{vdo.63,nag.75,mak.76,wol.77,vol.14}. For the gapped modes, the masses have the form,
\begin{equation}
M_{J^c} = a_J^c (T,P)\,\,\Delta(T,P),
\end{equation}
where $\Delta$ is the Bogoliubov fermion mass (the energy gap of the superfluid) and $a_{J}^c$ is a pressure and temperature-dependent numerical coefficient \cite{sau.17}. Of particular interest are the real and imaginary squashing modes with quantum numbers and bare masses respectively,
\begin{align}
{J}^c = 2^{+}, \,\,\,\,\,\,&M_{2^+} = \sqrt{\frac{8}{5}}\Delta, \label{eq:qnum1}\\
{J}^c = 2^{-}, \,\,\,\,\,\,&M_{2^-} = \sqrt{\frac{12}{5}}\Delta. \label{eq:qnum2}
\end{align}

The observed masses, however, have coefficients renormalized  from these bare values due to several effects including Fermi liquid interactions and higher-order pairing interactions \cite{sau.81a}. While the dominant pairing channel that gives rise to superfluidity in $^3$He is $p$-wave, a sub-dominant, attractive $f$-wave ($L=3$) interaction is predicted by ferromagnetic, spin-fluctuation mediated pairing \cite{fay.68} with interaction strength denoted by $x_3^{-1}$, where $x_3$ = ln $T_{c3}/T_{c}$ and $T_{c3}$ would be the superfluid transition temperature from $f$-wave pairing in the absence of $p$-wave pairing~\cite{sau.86}. Non-zero values for $x_3^{-1}$ and the Fermi liquid interaction parameters, $F_2^s$ and $F_2^a$ (respectively,  spin-symmetric and spin-antisymmetric Landau parameters), would lead to observable shifts in the mode masses. Prior comparisons between the observed and the theoretical values of the mode masses have allowed for these two effects as well as for strong-coupling corrections to the energy gap \cite{ser.83} which incorporate physics beyond the weak-coupling BCS theory \cite{hal.90,dav.06,col.13b}.

\begin{figure*}
\includegraphics[width=1\textwidth]{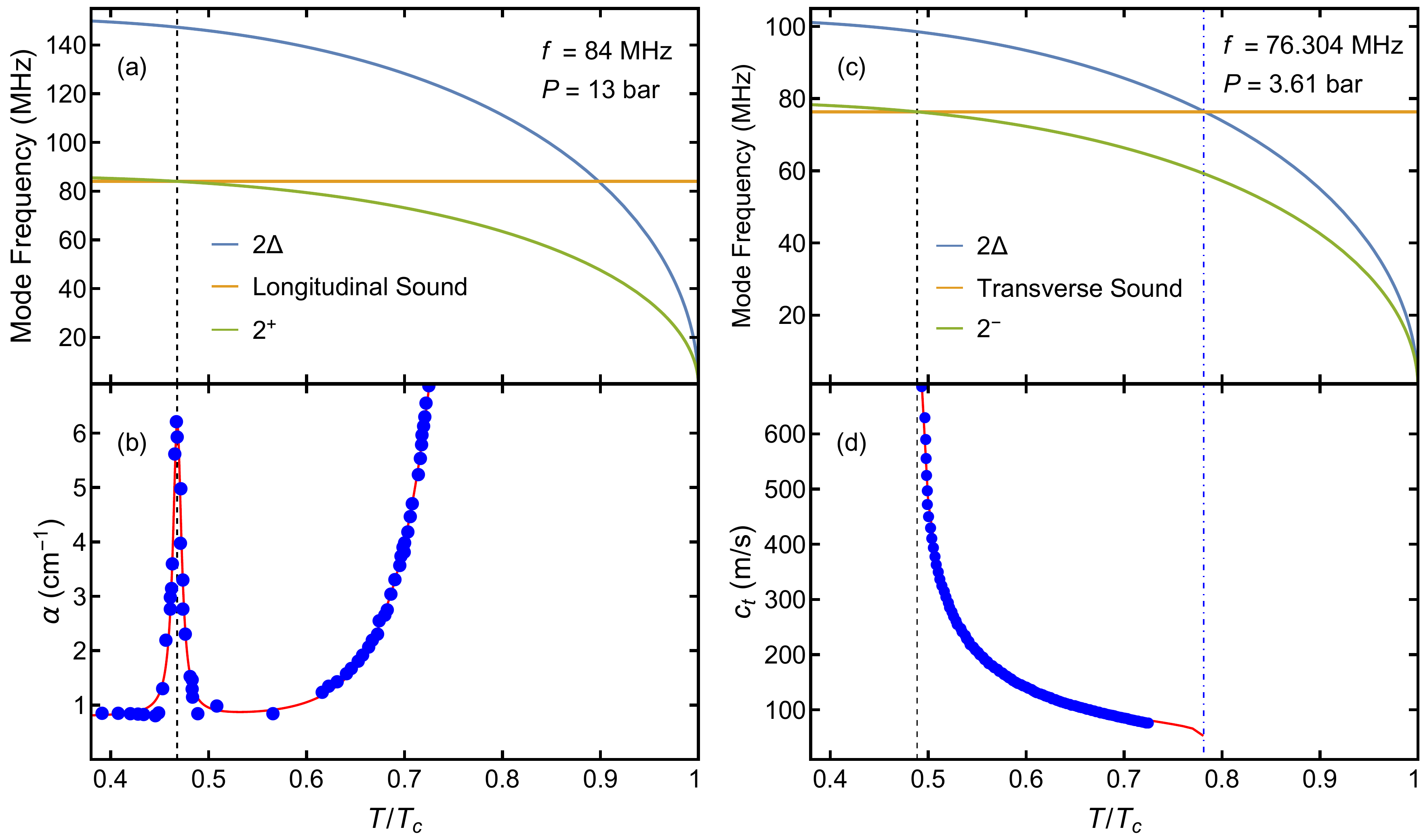}
\caption{Spectroscopic signatures of the $2^+$ mode at 13 bar and 84 MHz (left) and of the $2^-$ mode at 3.61 bar and 76.304 MHz (right). The blue trace in the top two panels shows the superfluid energy gap, $\Delta$, opening up as temperature is lowered while the green trace shows the corresponding increase in mode masses. For $\textbf{(a)}$ and $\textbf{(b)}$, longitudinal sound is excited at a fixed frequency of 84 MHz, represented by the orange trace. When the sound frequency crosses the mode energy, denoted by the vertical dashed line, a sharp resonance in the acoustic attenuation, $\alpha$, is observed, indicating the value of the $2^+$ mode mass, as seen in $\textbf{(b)}$ (blue circles are data from \cite{mas.80}, and the red line is a phenomenological model). For $\textbf{(c)}$ and $\textbf{(d)}$, transverse sound is excited at a fixed frequency of 76.304 MHz. Transverse sound can only propagate in the region between the black dashed and blue dot-dashed lines, where its frequency is between 2$\Delta$ and the $2^-$ mode. At the crossing, the transverse phase velocity, $c_t$, diverges indicating the value of the $2^-$ mode mass (blue circles are taken from data \cite{dav.08c}; the red line is from the theory \cite{moo.93}.)}
\label{fig:exp}
\end{figure*}

However, there are also strong-coupling corrections to the coefficients $a_{J}^c$, referred to as non-trivial strong-coupling corrections. The existence of these corrections has been noted in the literature \cite{koc.81,sau.81a} but a complete, dynamical theory of all Fermi liquid, $f$-wave, and strong-coupling effects has not yet been achieved. Koch and W$\ddot{\mathrm{o}}$lfle noted \cite{koc.81} that these non-trivial corrections to $a_{J}^c$ can be estimated using the strong-coupling corrections to the $\beta$-parameters of the Ginzburg-Landau (GL)  functional. The five $\beta$-parameters, $\beta_i$, are the coefficients of the fourth-order invariants of the order parameter~\cite{mer.73,sau.17}. The addition of strong-coupling corrections to $\beta_i$ and therefore $a_{J}^c$ is made possible by recent advances in determining the strong-coupling interactions and their temperature dependence~\cite{cho.07,wim.18}.

Here we apply a simple procedure along these lines that incorporates mass renormalization due to Fermi liquid, $f$-wave, and strong-coupling effects to both $\Delta$ and $a_{J}^c$. 
In terms of $\beta$-parameters, $a_J^c$ is given by \cite{sau.17} 
\begin{align}
a_2^+ &= 2\left(\frac{\frac{1}{3}(\beta_3+\beta_4+\beta_5)}{\beta_B}\right)^{1/2}\label{eq:bm1}\\
a_2^- &= 2\left(\frac{-\beta_1}{\beta_B}\right)^{1/2}\label{eq:bm2},
\end{align}
with $\beta_B$\,=\,$\beta_1 + \beta_2 + \frac{1}{3}(\beta_3+\beta_4+\beta_5).$ In the weak-coupling limit, the $\beta$-parameters satisfy the relation,
\begin{equation}
-2\beta_1=\beta_2=\beta_3=\beta_4=-\beta_5\label{eq:wcb},
\end{equation}
which reduces $a_2^+$ to $\sqrt{\frac{8}{5}}$ and $a_2^-$ to $\sqrt{\frac{12}{5}}$. The relation in Eq. (\ref{eq:wcb}) no longer holds when strong-coupling corrections are included. However, Eqs. (\ref{eq:bm1}) and (\ref{eq:bm2}) are still valid and can be used in conjunction with the strong-coupling $\beta$-parameters. Using this procedure, we find improved agreement for the $f$-wave pairing strength determined from several independent experiments.

\section{II. Experiment}
Experimental signatures of the collective modes are observed with acoustic spectroscopy. Superfluid $^3$He is able to support propagating sound at low temperature with either longitudinal \cite{mak.74} or transverse polarization \cite{moo.93,lee.99}. Collective modes in the superfluid  couple to sound for which there are both longitudinal and transverse restoring forces.  The latter is a unique property of superfluid $^3$He. This coupling allows one to probe the collective mode spectra of the  2$^-$ and 2$^+$ modes. Both are acoustically active and have sharp spectroscopic signatures \cite{dav.06}. The phase velocity and attenuation of sound waves diverge sharply when the frequency of sound matches the energy of the mode as shown in Fig. \ref{fig:exp}, providing a determination of the mode mass. While longitudinal sound has been used to measure both modes, the $2^-$ mode has a much stronger coupling to longitudinal sound than does the $2^+$ mode. This leads to a broad resonance for $2^-$ while the $2^+$ mode is sharp and very well-defined. Therefore, longitudinal sound measurements are only suitable for precise measurements of the $2^+$ mode. Transverse sound, on the other hand, can only propagate due to an off-resonant coupling to the $2^-$ mode \cite{moo.93,lee.99}. When the transverse sound frequency is less than the energy of the $2^-$ mode, sound propagation ceases abruptly giving a clear indication of the mode mass. Temperature, pressure, and frequency sweeps have been performed by several groups to map out the energy of the modes throughout the entire superfluid phase diagram \cite{mas.80,gia.80,fra.89,dav.06}. 

For temperature and pressure sweeps, a piezo-electric transducer is driven either continuously or with pulsed excitations at one of its odd harmonics. This method was employed by Mast $\textit{et al.}$~\cite{mas.80}, Giannetta $\textit{et al.}$~\cite{gia.80}, and Davis $\textit{et al.}$~\cite{dav.06,dav.08c}. While this method can be used to obtain precise values for attenuation and sound velocity, the frequency range is restricted to discrete harmonics of the transducer. Complementary to this, Fraenkel $\textit{ et al.}$ \cite{fra.89} employed pulsed excitation of a non-resonant ultrasound transducer to perform frequency sweeps at fixed temperature and pressure. In this case the frequency was swept through the mode and a Lorentzian absorption spectrum was observed. The frequency of the maximum absorption corresponds to the mode mass. The frequency sweep method, however, cannot extract the absolute attenuation~\cite{fra.89}. 

Davis $\textit{et al.}$ used LCMN susceptibility thermometry precise to within 30\,$\mu$K  and can be calibrated using fixed points from the Greywall melting-curve temperature scale \cite{gre.86}. Giannetta $\textit{et al.}$ used the Helsinki-scale \cite{alv.80} which is referenced to a different superfluid transition temperature than the more precise Greywall scale. For our analysis, we rescaled the temperatures reported by \cite{gia.80} to the Greywall melting-curve scale. Fraenkel $\textit{et al.}$ expressed temperature dependence in terms of the normal-fluid fraction, $\rho_n/\rho$, which we converted to reduced temperature, $T/T_c$, by interpolating data from Ref. \cite{par.85}.

\section{III. strong-coupling and $f$-wave corrections}
The experimental results indicate that $a_2^+$ is 10-15$\%$ smaller (depending upon pressure) than the bare value of $\sqrt{\frac{8}{5}}$. On the other hand, $a_2^-$ is only 1-4$\%$ larger than its bare value of $\sqrt{\frac{12}{5}}$. Therefore, Fermi liquid, $f$-wave, and strong-coupling effects must be included to obtain a consistent understanding of the mode masses.

The mode masses, $M_{2^+}$ and $M_{2^-}$, including renormalization due to Fermi liquid and $f$-wave interactions, were first calculated by Sauls and Serene~\cite{sau.17,sau.81a} in the weak coupling limit. We combine their result with  Eqs. (\ref{eq:bm1}) and (\ref{eq:bm2}) to obtain,
\begin{multline}
0 \,\,= \,\,M_{2^+}^2 - \left[4\left(\frac{\frac{1}{3}(\beta_3+\beta_4+\beta_5)}{\beta_B}\right)\Delta^2\right] +\\ \lambda(M_{2^+},T)( M_{2^+}^2- 4 \Delta^2)\left(\frac{2}{25} F_2^a + x_3^{-1}\left(\frac{M_{2^+}}{2\Delta}\right)^2\right)\label{eq:renorm1}
\end{multline}
\begin{multline}
0 \,\,= \,\,M_{2^-}^2 - \left[4\left(\frac{-\beta_1}{\beta_B}\right)\Delta^2\right] +\\ \lambda(M_{2^-},T)( M_{2^-}^2- 4 \Delta^2)\left(\frac{3}{25} F_2^s + x_3^{-1}\left(\frac{M_{2^-}}{2\Delta}\right)^2\right),\label{eq:renorm2}
\end{multline}
where $\lambda(M_{J^c},T)$ is the frequency and temperature dependent superfluid response function, evaluated at the mode mass \cite{moo.93}. This expression can be used in both the weak and strong-coupling limits with appropriate values of the $\beta$-parameters inside the square brackets. The $f$-wave interaction parameter $x_3^{-1}$ can be determined using Eqs. (\ref{eq:renorm1}) and (\ref{eq:renorm2}) with  sufficiently precise measurements of the masses and Fermi liquid parameters. Independent measurements of these parameters give values for $F_2^s$ ranging from -0.2 at 0 bar to +0.6 at 30 bar while $F_2^a$ is negative for all pressures, ranging from -0.9 at 0 bar to -0.1 at 30 bar \cite{hal.90}. Positive values for Fermi liquid or $f$-wave interactions increase the mass while negative values decrease the mass (see Ref \cite{sau.17}, Fig. 3).

\begin{figure*}
\includegraphics[width=1\textwidth]{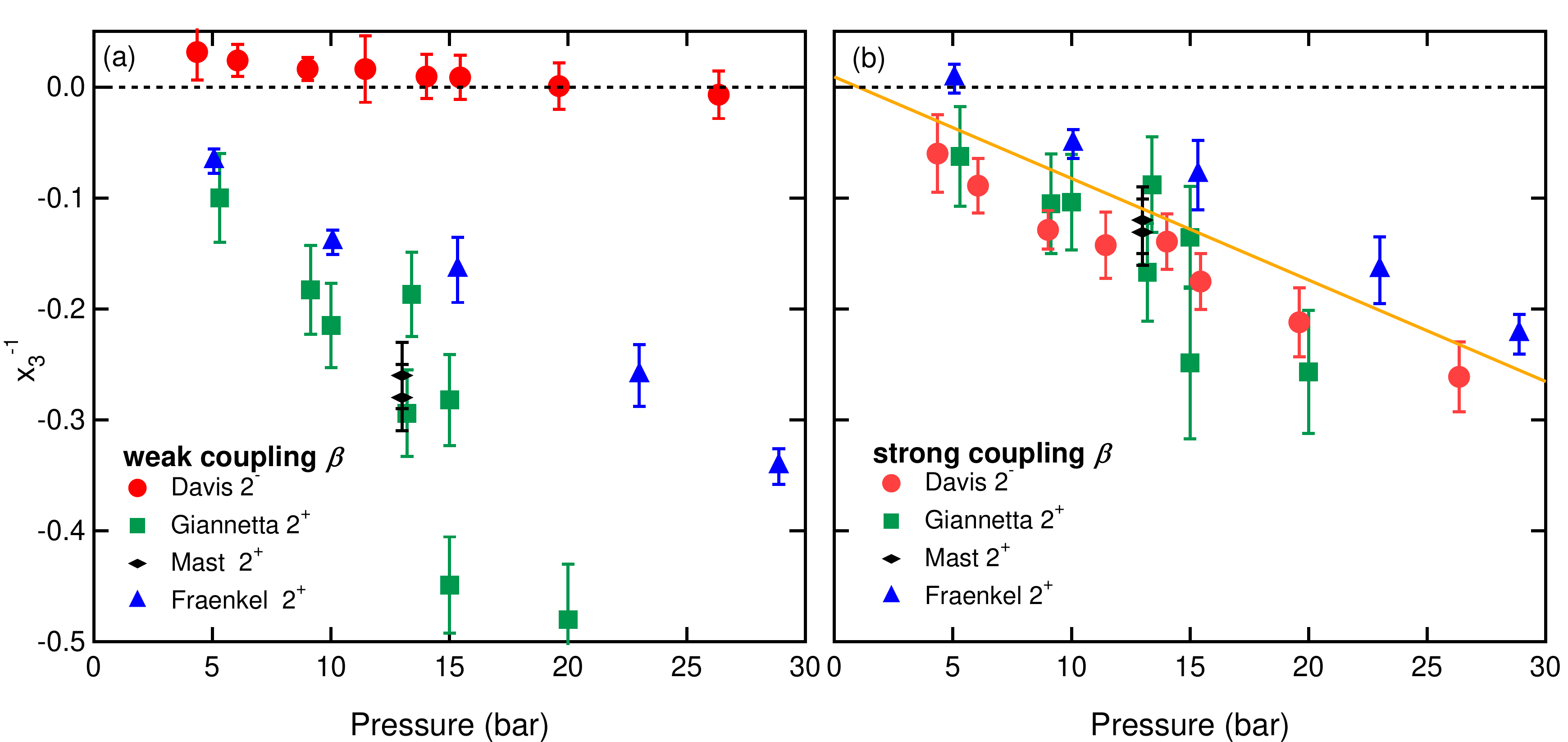}
\caption{$f$-wave interaction stength, $x_3^{-1}$, calculated using data from Davis $\textit{et al.}$ ~\cite{dav.08c}, Fraenkel $\textit{et al.}$~\cite{fra.89}, Giannetta $\textit{et al.}$~\cite{gia.80}, and Mast $\textit{et al.}$~\cite{mas.80}. The results in the left panel were calculated from Eqs (\ref{eq:renorm1}) and (\ref{eq:renorm2}) using the weak-coupling limits of $a_{J}^c$. The right panel incorporates strong-coupling corrections, which brings all four experiments into better agreement. A linear fit (orange solid line) through the entire dataset, weighted by uncertainties, yields a pressure dependence for the $f$-wave interaction strength reasonably consistent with all experiments.}
\label{fig:x3}
\end{figure*}

The gap used in our calculation is determined from Rainer and Serene's weak-coupling-plus model that extends the BCS theory to include strong-coupling interactions \cite{hal.90,rai.76}. This model is believed to accurately represent the energy gap, limited by the accuracy of measurements of the heat capacity jump \cite{hal.90}. Masuhara $\textit{et al.}$ \cite{mas.00} performed direct measurements of the energy gap using acoustic spectroscopy where they concluded that the weak-coupling-plus model overestimates the energy gap by $2-4\%$. However, what they believed to be the 2$\Delta$ pair-breaking edge was likely another collective mode with mass just below 2$\Delta$ \cite{dav.08b}. This new mode has mass between 1.97 and 1.99\,$\Delta$, which if incorrectly interpreted as the 2$\Delta$ pair-breaking edge, could lead to an erroneous conclusion. Other determinations of the gap have been performed by measuring quasi-particle damping \cite{tod.02}. The Fermi liquid parameters, $F_2^{a,s}$ have uncertainties discussed elsewhere~\cite{hal.90}. Accuracy of the temperature scale, the value of the energy gap, and the Fermi liquid parameters are the dominant sources of experimental uncertainty in our analysis.

Davis $\textit{et al.}$ calculated $x_3^{-1}$ in the weak-coupling limit from their transverse acoustic measurement of the $2^-$ mode mass. In this limit, the term in square brackets for Eq. (\ref{eq:renorm2}) reduces to $\frac{12}{5}\Delta^2$ and the results are shown in Fig. \ref{fig:x3}(a).  To make a comparison with the $2^+$ mode, we have used the data from Fraenkel $\textit{et al.}$, Giannetta $\textit{et al.}$, and Mast $\textit{et al.}$ to calculate $x_3^{-1}$ in the weak-coupling limit, where the term in square brackets for Eq. (\ref{eq:renorm1}) reduces to $\frac{8}{5}\Delta^2$. As seen in Fig. \ref{fig:x3}(a), there is significant disagreement between the $x_3^{-1}$ inferred from these 4 experiments. We find that this discrepancy is removed by incorporating strong-coupling corrections to the coefficients $a_{J}^c$.

Strong-coupling corrections lead to a pressure and temperature dependence for the terms in square brackets in Eqs. (\ref{eq:renorm1}) and (\ref{eq:renorm2}). The pressure dependences of the $\beta_i$ have been calculated by Wiman \cite{wim.18} using a microscopic model of quasiparticle scattering along with normal state Fermi-liquid data and measurements of the specific heat jump at $T_c$. Independently, Choi $\textit{et al.}$ \cite{cho.07} calculated the pressure dependence of $\beta_i$ from measurements in superfluid $^3$He. Wiman \textit{et al.} \cite{wim.18,wim.15} showed that the temperature dependence of the $\beta_i$ can be taken to be proportional to $T/T_c$ allowing them to extend the applicability of the GL theory to lower temperatures, based in part on Serene and Rainer's strong-coupling theory \cite{ser.83}. The pressure and temperature dependent strong-coupling $\beta$-parameters are given by 
\begin{equation}
\beta_i (T,P) = \beta_i^{W.C.} + \frac{T}{T_c} \Delta\beta_i(P)
\label{eq:betasc}
\end{equation}
where $\beta_i^{W.C.}$ is the weak-coupling value and $\Delta\beta_i(P)$ is the pressure dependent deviation away from the weak-coupling values at $T_c$ reported in \cite{cho.07}. Temperature scaling of the $\beta$-parameters has also been used in studies of superfluid $^3$He in silica aerogel \cite{cho.07,ger.02a,sha.03}. This linear temperature scaling weakens strong-coupling effects as temperature is lowered.  However, at sufficiently low temperatures, estimated to be approximately 0.3 $T_c$,  the linear dependence is expected to break down \cite{Sau.18}. The overall uncertainty in our analysis is indicated by error bars.

With the strong-coupling corrections that have been presented here, $a_2^+$ decreases with pressure while $a_2^-$ increases. When combined with the $F_2^a$ and $F_2^s$ corrections, we obtain a consistent determination of $x_3^{-1}$ from independent experiments on two different order parameter collective modes, as seen in Fig. \ref{fig:x3}(b). The left panel shows the $x_3^{-1}$ determinations using the weak-coupling values for $a_{J}^{c}$ while the right hand panel uses the strong-coupling $\beta_i$, bringing the four experiments into better agreement. While the $f$-wave interaction parameter only has a pressure dependence, its effect on the masses also has a temperature dependence inherited from the superfluid response function, $\lambda(M,T)$. This implicit dependence leads to measurements at the same pressure being shifted by different amounts.

To within experimental uncertainty, the four experiments find $x_3^{-1}$ varying consistently from close to zero at low pressures to -0.25 at high pressures.  A linear fit through the entire dataset, weighted by uncertainties, yields a pressure dependence for $x_3^{-1}$, 
\begin{equation}
x_3^{-1} (P) = 0.0091 -\, 0.0092 \,\,P/(\mathrm{bar}).
\label{eq:x3press}
\end{equation}
The $f$-wave interaction parameter is negative throughout the phase diagram, indicating an attractive pairing interaction in this channel, consistent with pairing mediated by ferromagnetic spin-fluctuations.  The magnitude of $x_3^{-1}$ can be used to calculate the instability temperature at which liquid $^3$He would undergo a superfluid transition with $f$-wave Cooper pairs, if the $p$-wave channel did not exist. For the present values of $x_3^{-1}$ at high pressure, this temperature is 90 $\mu$K at 34 bar. It is also noteworthy that $x_3^{-1}$ is close to zero at zero pressure. 

Superfluid $^3$He and superconductors are usually investigated assuming a single pairing channel. However, the possibility of pairing in multiple angular momentum channels has been predicted to exist in certain high-temperature superconductors \cite{rai.98, wan.99}. Here, we find evidence that the dynamics of superfluid $^3$He are indeed best modeled by a pairing potential with multiple angular momentum channels.

\section{IV. Conclusion}

The new strong-coupling analysis of the Higgs masses of the $J=2$ collective modes in superfluid $^3$He are significantly different from the earlier work which only accounts for strong-coupling corrections to the energy gap. Our work incorporates non-trivial strong-coupling using the pressure and temperature dependent $\beta$-parameters of the time dependent GL theory. We find that the measurements of the collective modes from transverse and longitudinal acoustics are sufficiently accurate that it is possible to extract the pairing interaction in the $f$-wave channel. Observations from two different collective modes, $J^c = 2^+$ and $2^-$, indicate that  $f$-wave pairing is attractive and is stronger with increasing pressure, consistent with spin-fluctuation mediated pairing.  The consistency between the results suggests that the theory of the collective modes, and correspondingly the Higgs masses, is now well-established.

\section{Acknowledgements}
We are grateful to J. A. Sauls and J. J. Wiman for useful discussions regarding the strong-coupling corrections to the $\beta$-parameters. We are also thankful to J. M. Parpia for guidance in converting $\rho_n/\rho$ to temperature. This work was supported by the National Science Foundation, Division of Materials Research (Grant No. DMR-1602542).
\bibliography{Manref}

\end{document}